%Paper: 9202016
%From: Malcolm Perry (on john.amtp.cam.ac.uk) <malcolm@amtp.cam.ac.uk>
%Date: Wed, 5 Feb 92 17:22:34 GMT

%
% Use plain TeX for this paper.
%

\magnification = 1200
\font\cs=cmr10 scaled \magstep4
\rightline{DAMTP R92-8}\vskip 1cm
\cs
\centerline{Path Integrals and Perturbative Expansions }\vskip 0.2cm
\centerline{for}\vskip 0.2cm
\centerline{Non-Compact Symmetric Spaces}
\tenrm
\vskip 1.5 cm
\centerline{Noah Linden}
\centerline{and}
\centerline{Malcolm J. Perry}
\bigskip
\centerline{Department of Applied Mathematics and
Theoretical Physics,} \centerline{University of
Cambridge,} \centerline{Silver Street,}
\centerline{Cambridge,}
\centerline{CB3 9EW,}
\centerline{England.}
\vskip 1.5 cm
\centerline{January 1992}
\vskip 1.5 cm
\bf Abstract: \rm
We show how to construct path integrals for  quantum mechanical
systems where the space of configurations is a general
non-compact symmetric space. Associated with this path integral
is a perturbation theory which respects the global structure
of the system. This perturbation expansion is evaluated for
a simple example and leads to a new exactly soluble model.
This work is  a step
towards  the construction  of a strong coupling perturbation theory for
quantum gravity.

\vskip 2 cm
\leftline{{\bf E-mail:} malcolm@amtp.cam.ac.uk,}
\leftline{\hskip 1.43cm nl101@phx.cam.ac.uk}

\vfill
\eject
\def\Complex{{\bf C}}
\def\GL{GL(n,\Real)}
\def\half{{1\over 2}}
\def\ltimes{\widetilde\times}
\def\Real{{\bf R}}
\def\vvec{{\bf v}}

\def\svec{{\bf s}}
\def\Symmp{Symm^+(n,\Real)}
\def\uvec{{\bf u}}
\newcount\section
\section=0
\def\sec{\the\section}
\newcount\equationnumber
\def\clearequationnumber{\equationnumber=0}
\def\eqn{\global\advance\equationnumber by 1
        \the\equationnumber }
\newcount\notenumber
\notenumber=0
\def\note{\global\advance\notenumber by 1
              \footnote{$^{\the\notenumber}$}}
\def\eqnu{\eqno(\sec.\eqn)}
\openup 1\jot
\line{\bf 1. Introduction\hfil}
\bigskip
\advance\section by 1
\clearequationnumber
The construction of a quantum theory of fundamental processes that is central
to physics. Despite this, quantization of even the most elementary
systems has proved to be far from straightforward. For a system whose
configuration space is the whole of the real line, \Real,
or any finite dimensional vector space, there are a number of
equivalent quantization schemes.
One approach is the canonical operator formalism in which the
phase space variables $x$ and $p$ are replaced by operators
$\hat x$ and $\hat p$, and their
Poisson bracket
$$ \{x,p\}=1 \eqnu $$
is replaced by the commutator
$$ [\hat x, \hat p] = i, \eqnu $$
the Heisenberg-Weyl algebra. Quantization is completed by
finding an irreducible unitary representation of this algebra;
it may be shown that any such representation is unitarily equivalent
to the usual one on the space of square-integrable complex-valued functions
on \Real,
 $L^2(\Real,\Complex)$ [1].

An alternative and very beautiful formulation of quantum mechanics on
\Real~
is the path integral. In this  approach, the transition amplitude
between an initial state $i$ at time $t_i$, and a final
state $f$ at time $t_f$, is
$$ G(f,t_f;i,t_i) = \int {\cal D}[x(t),p(t)]
e^{i\int_{t_i}^{t_f} dt ( p\dot x - H(p,x))} \eqnu $$
In the path integral, ${\cal D}[x(t),p(t)] $ is the Feynman-Wiener
measure on the space of all paths in phase space that interpolate
between the inital and final configurations. This formula applies
to any system with Hamiltonian $H$.
and  can be shown to be equivalent to the operator formulation
under certain circumstances [2].

In order to treat more complicated problems, we would like to extend
the ideas of path integration to more general situations such as are
encountered in condensed matter physics, gravitation and
$\sigma$-models. Our intention is to construct a path integral whose
perturbation theory  respects the global structure of the system.
 In a previous publication [3], we constructed a
quantum mechanical path integral for a system whose configuration space
is $\Real^+$,  the positive real line, and applied it to a
minisuperspace model of   quantum gravity.  We  showed how
the fact that the configuration space variable was restricted to be
positive led to a different form of the path integral measure than is
usually assumed, and that the unitarity of time evolution and the
question of topology change are  crucially affected by the correct
identification of the path integral  measure.

It  seems clear that related considerations will have a considerable
bearing on any complete quantum theory of gravity,
and this paper is a step in that direction. Specifically,
we extend the construction of the path integral in [3] to a more general
finite-dimensional non-compact configuration space of the form
$G/H$, where $G$ and $H$ are two Lie gorups.  As we shall show,
this path integral respects the global structure of the
system and allows the development of a perturbation theory which goes beyond
the conventional weak field expansion in analogous field theory systems.

Although we shall treat quantum mechanics in this paper, our motivation is
the field theoretic analog to which we intend to return in a future
paper.  Our primary aim here is to develop a  perturbation
scheme which does not have the drawbacks of the Feynman diagram approach
usually employed.  The first drawback is that conventional perturbation theory
is based on Gaussian integration, and this means that one allows the
integration variables to range from minus infinity to infinity which is  an
approximation that does not respect the global structure of general
non-linear systems.  Secondly, for many non-compact configuration spaces, the
normal coordinate expansion does not probe the whole space: for example the
matrix $\left(\matrix{-2&0\cr 0&-{1\over 2}\cr}\right) \in
SL(2,\Real)$ cannot be written in the form $\exp(\alpha^i T^i)$ where $T^i$ are
the generators of $SL(2,\Real)$.  Thus the Vilkovisky-DeWitt approach, while
reflecting the local geometry, does not allow one to probe the full range of
quantum fluctuations.

A further motivation comes from quantum gravity, for which the space of
configurations,  positive definite symmetric matrices at each point in
space modulo spatial diffeomorphisms, is an example of the field
theoretic extension of this work.  It is known that the weak field
perturbation theory for this system is unsatisfactory because it is
unrenormalizable,  however the approach adopted here is naturally
tailored to a strong field perturbation theory [4].
 Although this is the most interesting regime of
quantum gravity, it is the least understood and we expect the work described
here to be a step towards allowing us to probe this domain.

We should make it clear that the path integral described here is an
alternative to the usual configuration space or phase space path integrals
which are also correct in the sense that they satisfy the relevant
Schr\"odinger equation [2].  We argue, however, that the path integral
we propose is  useful if one wishes to do the sort of \lq\lq global"
perturbation theory we are interested in, whereas the configuration space path
integral is more adapted to weak field perturbation theory .

The construction we shall present takes as its starting point the canonical
quantum mechanics of the relevant system, which we shall take to be the finite
dimensional coset space $G/H$ where $G$ is non-compact and $H$ is compact.  As
has been explained elsewhere [5], [6], the procedure for finding an algebra of
functions on the classical phase space closed under Poisson brackets and
then turning these functions into operators leads to the \lq\lq
canonical" group  of the semi-direct product form, $W\ltimes G$,
where $W$ is a
vector space in which $G$ acts linearly with an orbit $G/H$.

The space of quantum states corresponds to unitary representations of this
group. There may be many inequivalent quantizations of this system,
in contrast to the case of quantum mechanics on \Real. The inequivalent
quantizations of $G/H$ are
labelled by irreducible representations $\pi_\chi$ of $H$, with the
quantization corresponding to $\pi_\chi$  realised on the Hilbert space of
square-integrable functions taking their  values in the vector space ${\cal
H}_\chi$ carrying the (finite-dimensional) irreducible representation
$\pi_\chi$ of $H$, {\it ie.} the Hilbert space is $L^2(G/H,{\cal
H}_\chi)$,~[5], [6].  Amongst these quantisations is the \lq\lq usual\rq\rq\
one on $L^2(G/H,\Complex)$, corresponding to the trivial representation of $H$.

As a first example,
we shall concentrate on the  case
of $n\times n$ positive symmetric matrices, $Symm^+(n,\Real)$, that is
symmetric matrices
whose eigenvalues are all positive definite.  This space  is
isomorphic to the coset space $GL(n,\Real)/O(n)$ [7] since $\GL$ acts
transitively on $\Symmp$ by conjugation,
$$Y\mapsto g^TYg\quad Y\in\Symmp, g\in\GL\ ,\eqnu $$
with little group at the identity, $O(n)$.
The canonical group [7] is $Symm(n,\Real)\ltimes\GL$ with
Lie algebra relations
$$\eqalign{ [\hat Y_{ab},\hat Y_{cd}] &= 0\cr
[\hat\pi_a{}^b,\hat\pi_c{}^d] &= i(\delta_a^d \hat\pi_c{}^b - \delta_c^b
\hat\pi_a{}^d)\cr
[\hat Y_{ab},\hat\pi_c{}^d] &= i(\delta_a^d \hat Y_{bc} + \delta_b^d
\hat Y_{ac}),\cr}\ \eqnu$$
where $Symm(n,\Real)$ is space of all real $n \times n$ symmetric matrices,
and is isomorphic to $\Real^{\half n(n+1)}$.
The natural $\GL$ invariant Hamiltonian on $\Symmp$ is
$$\hat H_0 = \half \hat\pi_a{}^b \hat\pi_b{}^a \ .\eqnu$$
In this paper, we shall treat only the quantisation in which ${\cal H}_\chi=
\Complex$ and in this case the representation of $\hat\pi$ and $\hat Y$ is
given by
$$\eqalign{
\hat\pi_a{}^c \psi(Y) &= -iY_{ab}\biggl(
{\partial\over\partial Y}\biggr)_{bc}\psi(Y)\cr
\hat Y_{ab}~\psi(Y) &= Y_{ab}~\psi(Y)\cr}\ .\eqnu$$
with
$$ \biggl({\partial\over\partial Y}\biggr)_{ab}
= (1+\delta_{ab}){\biggl({\partial\over\partial
 Y_{ab}}\biggr) \ \ \ \ {(no~~sum)}}  \eqnu. $$

The plan of this paper is as follows: in section 2 we outline how to carry out
Fourier analysis on $Symm^+(n,\Real)$ which we then use to construct a
formal path integral.  In section 3 we shall show how this form of the path
integral, rather than the configuration space or phase space form, leads
naturally to the \lq\lq global"  perturbation theory;  we then calculate
the first order in perturbation theory for the quantum mechanics of a particle
moving on $\Symmp$ in a simple potential, in order to demonstrate that the
terms in the  perturbation expansion may be evaluated in detail. The form of
this expansion suggests that this highly nonlinear model has an exact
solution.
We verify that this is indeed the case. In section four, we show how this
construction can be generalized to more general non-compact symmetric spaces.
Finally, some conclusions are presented in section 5.

\bigskip
\line{\bf 2. Fourier-Helgason Transform and Path Integral\hfil}
\bigskip
\advance\section by 1
\clearequationnumber

The Fourier modes $ e^{ipx} $ on the real line $\Real$ have two important
properties.  The first is that they are \lq\lq complete" in the sense
that the Fourier transform $\tilde f(p)$ of a function $f(x)$, given by
$$\tilde f(p) = \int_{-\infty}^\infty f(x)\ e^{ipx}\ dx, \eqnu $$
may be inverted by the Fourier inversion formula
$$  f(x) ={1\over 2\pi} \int_{-\infty}^\infty \tilde f(p)\
e^{-ipx}\ dp, \eqnu $$
and secondly that  they are eigenfunctions
\note{Since $e^{ipx}$ is not normalizable with respect to the usual
$L^2$ inner product, it is not, strictly speaking an eigenfunction; this
problem may, however, be treated by rigged Hilbert space techniques [8].}
 of the invariant differential operators on
$\Real$, the most important example of which is the Laplacian
${d^2\over dx^2}$.

These two properties are extremely useful since they allow one to
construct the propagator for a free particle, that is the propagator for the
Schr\"odinger  equation with free Hamiltonian $-\half{d^2\over dx^2}$. It is
$$G_0
(x_f,t_f;x_i,t_i) = \int_{-\infty}^\infty dp\ e^{ip(x_f-x_i)}\
e^{{1\over 2}ip^2(t_f-t_i)}\ . \eqnu $$

These Fourier modes are also used in the  construction of the
path integral for a particle moving on $\Real$. In order to find the phase
space path integral expression for the propagator
$\langle x_f,t_f\mid x_i,t_i\rangle$ one inserts complete sets
of states $\mid x \rangle$ and $\mid p\rangle$ at $N$ intermediate
intervals.  $\mid x\rangle$ is an eigenstate of the operator $\hat x$
with eigenvalue $x$ and  $\mid p\rangle$ is an eigenstate
of $\hat p$ with eigenvalue $p$. They satisfy the completeness
relations $$\int_{-\infty}^\infty dx \mid x\rangle\langle x\mid =
{1\over 2\pi}\int_{-\infty}^\infty dp \mid p\rangle\langle p\mid = 1 \ .
\eqnu $$
Thus
$$\eqalign{&\langle x_f,t_f|x_i,t_i\rangle\cr
&=\left(\prod_{r=1}^N \int_{-\infty}^\infty {dx_r}
\left({1 \over 2\pi}\right)^N
\int_{-\infty}^\infty {dP_r}\right)\
\langle x_f,t_f|x_N,t_N\rangle\langle x_N,t_N|P_{N},t_{N}\rangle\ldots\cr
&\ldots|x_{r+1},t_{r+1}\rangle\langle
x_{r+1},t_{r+1}|P_{r+1},t_{r+1}\rangle\langle P_{r+1},t_{r+1}|
x_{r},t_{r}\rangle\langle x_{r},t_{r}|\ldots \cr &\ldots\langle P_{1},t_{1}|
x_i,t_i\rangle.\cr}\eqnu$$
 We may now use the fact that
$$\langle x_{r},t_{r}|P_{r},t_{r}\rangle =
e^{iP_r{x_r}}\eqnu$$ and
$$\langle P_{r+1},t_{r+1}|
x_{r},t_{r}\rangle = e^{ -iP_{r+1}{x_{r}} - i(t_{r+1}-t_r)H(x_r,P_{r+1})}
\eqnu$$
to order $(t_{r+1}-t_r)$ and hence implicitly to lowest order in $\hbar$.

Therefore
$$\eqalign{\langle x_f,t_f|x_i,t_i\rangle& \simeq \lim_{N \to
\infty} \left(\prod_{r=1}^N \int_{-\infty}^\infty {dx_r}
\left({1 \over 2\pi}\right)^N
\int_{-\infty}^\infty {dP_r}\right)\ \delta(x_f-x_N)\cr
&\exp i\Biggl\{\sum_{r=0}^{N-1}\biggl(P_{r+1} ({x_{r+1}-x_r}) - \Delta t~
H(x_r,P_{r+1})\biggr)\Biggr\}\cr}
\eqnu$$
so that, were the limit to exist,
$$\langle x_f,t_f|x_i,t_i\rangle~ \hbox{\lq} =\hbox{\rq}\
Z\int {\cal D}P
 \int {{\cal D}x} \exp \Biggl\{i\int_{t_i}^{t_f} dt \left(P (t) {\dot x(t)} -
H(x,P)\right)\Biggr\}\eqnu$$
where $Z$ is a normalisation constant.

This last equation is the usual starting point for perturbative calculations.
If $H=H_0 + \lambda V$ with some Hamiltonian $H_0$ for which the path
integral is known, and potential $\lambda V$, then by standard methods one
can calculate the path integral perturbatively as a series in positive powers
of $\lambda$. However, since the limit $N\to\infty$ does not really exist, one
can argue that the path integral is largely  a mnemonic
 and is correct in as much as it reproduces this expansion correctly.
This is not to say that the path integral is devoid of content;
indeed the path integral can be used to treat various phenomena
non-perturbatively,  an example is provided by the use of instantons
to treat tunneling processes.
 The expansion of the
propagator $G_V$ for $H=H_0+\lambda V$ in positive powers of $\lambda$
is most
easily derived from its integral equation [9], namely
$$G_V(x_f,t_f;x_i,t_i) = G_0(x_f,t_f;x_i,t_i) - \lambda \int_{t_i}^{t_f}dt
\int_{-\infty}^{\infty}
dx~~G_0(x_f,t_f;x,t) V(x) G_V(x,t;x_i,t_i)\ . \eqnu$$
Thus the perturbation series may be written, schematically, as
$$G_V = G_0 +  \lambda G_0VG_0 + \lambda^2G_0VG_0VG_0 + ....
,\eqnu$$ where
$$G_0VG_0 =  \int_{t_i}^{t_f}dt\int_{-\infty}^{\infty} dx G_0(x_f,t_f;x,t)
V(x)
G_0(x,t;x_i,t_i)\ ,\eqnu$$
and so forth for higher order terms.

We have set out this well-known derivation in detail since it may be repeated
in much more general situations such as the general (non-trivially
induced) quantisations of motion on the arbitrary finite-dimensional coset
spaces $G/H$ in which we are interested.

Rather than treating the general case here, we shall start
with the example of motion on the space of $n\times n$ positive
symmetric matrices $\Symmp$.  The beginning
of this section follows Terras [10] closely.
The analogs of the Fourier modes in this case are the so-called power functions
$p_{\svec}(Y[k])$, where $Y\in\Symmp$. These functions are labelled
by $\svec\in \Complex^n$, $k\in O(n)$
and $Y[k]$  means conjugated $Y$ by $k$, thus $Y[k]=k^TYk$. Then
$$ p_{\svec}(Y) = \prod_{j=1}^n |det\ Y_j|^{s_j}\ , \eqnu$$
where
$Y_j$ is the $j\times j$ matrix obtained by taking the first $j$ rows and
columns of $Y$.  Power functions  are eigenfunctions of the invariant
differential operators $L$. For $GL(n,\Real)$, a basis for the commutative
ring of invariant differential operators is given by
$$ L^{(r)} = {\rm Tr}~\pi^r \eqnu $$
with $r=1\ldots n$. For any invariant differential operator $L$,
the eigenvalues $\lambda_L$ depend only on
$\svec$ $$L\ p_\svec (Y[k]) = \lambda_L(\svec)~ p_\svec (Y[k])\
.\eqnu$$

The Fourier-Helgason transform $\tilde f(\svec,k)$, of a function $f(Y)$ on
$\Symmp$ is given by
$$\tilde f(\svec,k) = \int_{\Symmp} d\mu(Y)~f(Y)~\overline{ p_\svec (Y[k])}
\eqnu$$
 where $ d\mu(Y)$ is the  measure on $\Symmp$ invariant under
$\GL$, and the bar indicates complex conjugation.  The inversion formula is
$$f(Y) = \omega_n \int_{ \svec=\vvec+i\uvec}ds
\int_{\bar k \in K/M}d\bar k
\ \tilde f(\svec,k)~p_s(Y[k])~|c_n(\svec)|^{-2}\eqnu$$
where
$$\vvec = ({1\over 2},{1\over 2} .....,{1\over 2},{1\over 4}
(1-n)) \eqnu $$
and each component of $\uvec$ is real and runs from $-\infty$ to $\infty$.
$$\omega_n = \prod_{j=1}^n {\Gamma(j/2) \over j (2\pi i)\pi^{j/2}}
\eqnu$$
and the Harish-Chandra $c$-function is given in terms of the beta-function
$$ B(x,y) = {\Gamma(x)\Gamma(y)\over \Gamma(x+y)}\ ,\eqnu$$
as
$$c_n(\svec) = \prod_{1\leq i\leq j \leq n-1} {B(\half, s_i +....+ s_j +
\half(j-i+1)) \over B(\half,\half(j-i+1))}\ .\eqnu$$
Finally, the space over which the $k$'s are integrated is the coset space
$K/M$
where $K$ is $O(n)$ and $M$ is the set
of diagonal matrices with non-zero entries of $\pm 1$. That this integral is
over $\bar K$ rather than $K$ arises because
of the degeneracy in specifying $k$ given a particular power function coming
from a specified $Y$. In other words $M$ is the
automorphism group of the invariant power functions. By convention, one
normalizes the $\bar k$ integration over $K/M$ to be unity
$$ \int_{K/M} d\bar k = 1 \eqnu$$

For simplicity we shall
write the above inverse transform as
$$f(Y) = \int d\mu_{Helg}(\svec,k)
\ \tilde f(\svec,k) p_\svec(Y[k])  \eqnu$$
As in the case of quantum mechanics on $\Real$, we may use this transform to
give an expression for the free propagator where the Hamiltonian is the
invariant Laplacian on $\Symmp$.
The propagator $G_0(Y_f,t_f;Y_i,t_i)$ is the  solution to
$$ H_0 G_0(Y_f,t_f;Y_i,t_i) = i {\partial\over\partial t_i}
G_0(Y_f,t_f;Y_i,t_i)
,\eqnu$$
the Schr\" odinger equation,
with $H_0 = \half \hat\pi_a{}^b \hat\pi_b{}^a$ and
$\lim_{t_i\to t_f}G_0(Y_f,t_f;Y_i,t_i)= \delta(Y_f,Y_i)$, and it is given by
$$ G_0(Y_f,t_f;Y_i,t_i) = \int d\mu_{Helg}(\svec,k)
 p_{\svec}(Y_i[k]) \overline{p_\svec(Y_f[k])}\ e^{i\lambda_{H_0}(\svec)
(t_f-t_i)}
.\eqnu$$ We may also use the power functions to form a path
integral for this system just as in the case of quantum mechanics on $\Real$.
We insert complete sets of states,
$$\int d\mu_{inv}(Y)~|Y\rangle\langle Y|  = \int d\mu_{Helg}(\svec,k)~
|\svec,k\rangle\langle\svec,k|=1
\eqnu$$
to produce the following expression for the propagator associated with the
Hamiltonian $H_0 + \lambda V$
$$\eqalign{ \langle Y_f,t_f|Y_i,t_i\rangle = &\int
 \left(
\prod_{n=1}^N d\mu_{Helg}(\svec_n,k_n) d\mu_{inv}(Y_n)
\right)
e^{\sum_{n=1}^N \ln \bigl({p_\svec(Y_n[k])
\overline{p_\svec(Y_{n-1}[k])}\bigr)}}\cr
&e^{i\sum_{n=1}^N(\lambda_{H_0}(\svec_n)+\lambda V(Y_n))\Delta t
}.\cr}\eqnu$$  We note that this is not the usual phase-space
path integral.  However, expanding in powers of $\lambda$ gives the usual
canonical perturbation expansion in which all integrals are over measures and
ranges appropriate to the global structure of the problem under consideration.
\bigskip
\line{\bf 3. The Perturbation Expansion\hfil}
\bigskip
\advance\section by 1
\clearequationnumber
We have argued that if we want to develop a perturbation
theory that respects the global structure of the space of configurations, we
should use (2.27) rather than the usual phase space path integral, which is
tailored to perturbation theory using Gaussian integrals and hence only
respects the local geometry.  The usual expression does, at least, have the
benefit of allowing calculation of the perturbation series. We now show that
the terms in the perturbation series based on (2.27) may also be calculated
even though at first sight the relevant integrals look intractable.

For definiteness we consider the case of $2\times 2$ positive symmetric
matrices
and the Hamiltonian $H_0 + \lambda \ln\det(Y)$, where $H_0$ is the free
Hamiltonian, given by the invariant Laplacian. We now evaluate $G_V$ to the
lowest order in perturbation theory in $\lambda$.

The first order term in the expansion is thus
$$\lambda \int_{t_i}^{t_f}dt\int d\mu_{inv}(Y)\ G_0(Y_f,t_f;Y,t)~\ln\det(Y)
\ G_0(Y,t;Y_i,t_i)\ .\eqnu$$
We  show that this is equal to
$$\left(\half (t_f-t_i)\left(\ln\det(Y_i) + \ln\det(Y_f)\right)
\right)G_0(Y_f,t_f;Y_i,t_i)\ .\eqnu$$
Thus with
$$Y =\left(\matrix{Y_{11}&Y_{12}\cr
               Y_{12}&Y_{22}\cr}\right)\ ,\eqnu$$
it follows that
$$\eqalign{H_0 &= \half \hat\pi_a{}^b \hat\pi_b{}^a\cr
           &= -\half\biggl[ \left( 2 Y_{11}{\partial\over \partial Y_{11}} +
Y_{12}{\partial\over \partial Y_{12}} \right)^2 +
 \left( 2 Y_{12}{\partial\over
\partial Y_{11}} + Y_{22}{\partial\over \partial Y_{12}} \right)
\left(  Y_{11}{\partial\over \partial Y_{12}} +
2Y_{12}{\partial\over \partial Y_{22}} \right) \cr
&\quad + \left(  Y_{11}{\partial\over \partial Y_{12}} +
2Y_{12}{\partial\over \partial Y_{22}} \right)
\left( 2 Y_{12}{\partial\over
\partial Y_{11}} + Y_{22}{\partial\over \partial Y_{12}} \right) +
\left( 2 Y_{22}{\partial\over \partial Y_{22}} +
Y_{12}{\partial\over \partial Y_{12}} \right)^2\biggr]
\ .\cr}\eqnu$$
Thus since the eigenfunctions of the Laplacian are
$$ p_\svec(Y) = Y_{11}^{s_1} (Y_{11}Y_{22}-Y_{12}^2)^{s_2}\eqnu$$
it follows that the spectrum of the Hamiltonian specified by
$$H_0 p_\svec(Y)  = \lambda_{H_0}(\svec)p_\svec(Y)\eqnu$$
is
$$\lambda_{H_0}(\svec) = -2s_1^2 -4s_2^2 - 4s_1s_2 -s_1\ . \eqnu$$
Thus the propagator is given by
$$ G_0(Y,t;Y_1,t_1) = \int d\svec \int d\bar k |c(\svec)|^{-2}
 p_\svec(Y[k]) \overline{p_\svec(Y_1[k])}\ e^{-i\lambda_{H_0}(\svec)(t-t_1)}
\eqnu$$
The Harish-Chandra $c$-function evaluated along the relevant contour is
$$|c(-\half + iu_1 ,{1\over 4}+iu_2)|^{-2} = \pi u_1\tanh \pi u_1\ .
\eqnu$$ Now, given the form of $d\mu_{Helg}$
$$ \eqalign{ 0 &= \int d\mu_{Helg}~{\partial\over\partial u_2}\biggl(
p_\svec(Y[k]) \overline{p_\svec(Y_1[k])}\
e^{-i\lambda_L(\svec)(t-t_1)}\biggr)\cr &=-i\int
d\mu_{Helg}~\biggl(\ln\det(Y_1) - \ln\det(Y) -4 i (t-t_1)(s_1 + 2 s_2
)\biggr)\cr &\quad\quad\times p_\svec(Y[k]) \overline{p_\svec(Y_1[k])}\
e^{-i\lambda_L(\svec)(t-t_1)}\ .\cr}
\eqnu$$
But along the contour used in the inverse Helgason transform,
$$s_2~\overline{p_\svec(Y_i[k])}=
(-{\partial\over\partial\ln\det(Y_i)}+\half)\overline{p_\svec(Y_i[k])}\ ,
\eqnu$$ and
$$\eqalign{ s_1\overline{p_\svec(Y_i[k])} = -(
(Y_i)_{11} {\partial\over\partial(Y_i)_{11}}  +
&(Y_i)_{22} {\partial\over\partial(Y_i)_{22}} +
(Y_i)_{12} {\partial\over\partial(Y_i)_{12}} \cr
&+1 -2{\partial\over\partial\ln\det(Y_i)}   )
\overline{p_\svec(Y_i[k])} \ ,\cr}
\eqnu$$
hence
$$ \eqalign{\ln\det(Y)\ G_0(Y,t;Y_i,t_i) = \biggl(\ln\det(Y_i)  +4 i
(t-t_i)(
&(Y_i)_{11} {\partial\over\partial(Y_i)_{11}}  +
(Y_i)_{22} {\partial\over\partial(Y_i)_{22}} +\cr
&(Y_i)_{12} {\partial\over\partial(Y_i)_{12}})\biggr)\ G_0(Y,t;Y_i,t_i)\
.\cr}\eqnu$$
This allows us to do the $Y$ integration  using the fact
that
 $$\int d\mu_{inv}(Y)~ G_0(Y_f,t_f;Y,t)G_0(Y,t;Y_i,t_i) =
G_0(Y_f,t_f;Y_i,t_i)\ . \eqnu$$
Hence we find that
$$\eqalign{G_1(Y_f,t_f;Y_i,t_i) = \lambda\int_{t_i}^{t_f}dt
&\biggl[(\ln\det(Y_i)  + 4i
(t-t_i)\biggl(
(Y_i)_{11} {\partial\over\partial(Y_i)_{11}}  +
(Y_i)_{22} {\partial\over\partial(Y_i)_{22}} +\cr
&(Y_i)_{12} {\partial\over\partial(Y_i)_{12}}\biggr] \biggr)
G_0(Y_f,t_f;Y_i,t_i)\ ,\cr}
\eqnu$$
leading to
$$G_1(Y_f,t_f;Y_i,t_i) = \lambda\left(\half (t_f-t_i)\bigl(\ln\det(Y_i) +
\ln\det(Y_f) \bigr)\right)G_0(Y_f,t_f;Y_i,t_i)\ ,\eqnu$$
as in eqn (3.2).

It may be seen that the techniques used in this example allow the calculation
of the perturbation expansion for a large class of interactions, leading to
expressions of the form
$$G_n(Y_f,t_f;Y_i,t_i) = \lambda^n
f_n(Y_f,t_f;Y_i,t_i)\ G_0(Y_f,t_f;Y_i,t_i)\ .\eqnu$$
where $f_n$ is a multiplicative function.

As a last point in this section, we remark that, our expression for the first
order term for the perturbation expansion for motion on positive symmetric
$2\times 2$ matrices in the potential $V(Y) = \lambda\ln\det(Y)$ suggests the
following Ansatz for the exact propagator for this system:
$$G_V(Y_f,t_f;Y_i,t_i) = G_0(Y_f,t_f;Y_i,t_i) e^{\left(
\half\lambda(t_f-t_i)\left(\ln\det(Y_i)+\ln\det(Y_f)\right) -
{1\over 24}\lambda^2(t_f-t_i)^3\right)}\ .\eqnu$$
That this is the correct exact expression may be checked by direct substitution
into the relevant Schr\"odinger equation.  We have thus added to the relatively
small number of quantum systems whose  propagator may be found in closed form.
\note{We believe that there is an error in Feynman and Hibbs [9] p64 equation
(3-62) and that the correct last term in that equation should be $-{f^2T^3\over
24m}$ rather than $-{fT^3\over 24}$}
\bigskip
\line{\bf 4. General Non-Compact $G/H$.\hfil}
\bigskip
\advance\section by 1
\clearequationnumber

The same methods that were applied to quantum mechanics on positive symmetric
matrices can be applied directly to the general case of non-compact $G/H$,
at least for the trivially induced representation of $H$,
and provided that $H$ is a maximal compact subgroup of $G$.
The ingredients of this construction are a knowledge of the invariant
Laplacian $L^{(2)}$, its eigenfunctions, the Helgason transform
and its inverse. These building blocks may be found in
explicit detail in Helgason [11], Barut and Ra\c czka [12] and Terras [10].
The result is essentially the same as equation 2.27, namely
$$
\eqalign{
\langle Y_f,t_f \mid Y_i,t_i \rangle &=
\int \Biggl( \prod_{n=1}^N d\mu_{Helg}(\svec_n,k_n) d\mu_{inv}(Y_n)\Biggr)
e^{\sum_{n=1}^N \ln(p_\svec(Y_n[k])\overline{p_\svec(Y_{n-1}[k])})}\cr&
e^{i\sum_{n=1}^N (\lambda_{H_0}(\svec_n) + \lambda V(Y_n)) \Delta t}\cr}
\eqnu $$
This is the transition amplitude between the initial state
described by an element of $G/H$ given by $Y_i$, and a final state
given by $Y_f$. As before, $Y[k]$ is $Y\in G/H$ conjugated by $k\in H$,
and $p_\svec(Y)$ are the power functions associated with the Laplacian
$L^{(2)}$, which is taken to be a multiple of the free Hamiltonian
$H_0$ with spectrum $\lambda_{H_0}$. In the path integral itself,
$d\mu_{inv}$ is the canonical invariant measure on $G/H$, that is the
volume form derived from the same metric that determines $L^{(2)}$.
$d\mu_{Helg}$ is the measure for the inverse Helgason transform. This
expression is quite complicated, but Gindikin and Karpelevic [13]
have given an explicit form for the measure for such $G/H$.
We do not record the specific forms here, but refer to original literature
on this topic.

\bigskip
\line{\bf 5. Conclusion\hfil}
\bigskip
\advance\section by 1
\clearequationnumber
We have constructed a path integral for quantum mechanics on the space of
positive symmetric matrices whose natural perturbation theory respects the
global structure of the system.  A more usual approach to the problem would be
to use the phase-space or configuration space path integral.
  In the former case, one has (schematically)
$$ G = \int d\mu_{Liouville} e^{i\int~dt~(p\dot q - H)}\ ,\eqnu$$
and it can be shown [2] that this expression satisfies the Schr\"odinger
equation, so it must be equal to our expression.  However, although the above
equation is correct it is not clear how to use it to develop the sort of
expression we are interested in.

An alternative way to treat many non-linear systems using a phase-space path
integral is to follow Faddeev [14] and embed such a system in a vector space
(for which our expression becomes the usual one) and use the constraints to
restrict integration to the submanifold of interest.  Unfortunately, this
avenue is not open to us in the class of examples we have treated here since
$\Symmp$, rather than being defined by an equality (eg. a 2-sphere is defined
by $x^2+y^2+z^2= 1$ in $\Real^3$) is defined by inequalities (eg. the one
dimensional case is just the positive real line $x>0$).

A further way one might have tried to treat the system dealt with
in section three
would be to realise
that every positive symmetric matrix is the exponential of an arbitrary
symmetric matrix, and that the exponential map is one-to-one and onto, so that
we could have equivalently treated the space of arbitrary symmetric matrices
which is a vector space.  Apart from the fact that in a field theoretic
version of this problem such a transformation would be extremely delicate
(even $e^{\phi(x)}$ is ill-defined for usual scalar field theory), this
isomorphism is not helpful in the case at hand.  The reason is that, although
the space of $n\times n$ symmetric matrices is isomorphic to $\Real^{\half
n(n+1)}$ topologically, the natural metric on the  former (which is the one we
are using) has constant negative curvature.  Furthermore the Hamiltonian on
$\Real^{\half n(n+1)}$ in which we are interested  is the one invariant under
$\GL$ and so rather than the usual Laplacian   $-\sum{d^2\over dx_i^2}$
on $\Real^{\half n(n+1)}$, we
need a quantisation adapted to this Hamiltonian.  In other words, in order to
treat the Hamiltonian of interest, we have to use a \lq\lq strange\rq\rq\
quantisation of $\Real^{\half n(n+1)}$ using $Symm(n,\Real)\ltimes\GL$ rather
than the Heisenberg-Weyl group.  In fact, viewed as a theory on $\Real^{\half
n(n+1)}$, our system for the free Hamiltonian is related to one of the
integrable quantum systems discussed for example by Olshanetsky and
Perelomov[15].

Lastly we would like to put forward a path integral for a field theory taking
its values in positive symmetric matrices.  By analogy with the quantum
mechanical case, we suggest that the relevant expression should be
$$\eqalign{ G = &\prod_x\int
 \left(
 d\mu_{Helg}(\svec(x),k(x)) d\mu_{inv}(Y(x))
\right)
e^{\int dx \ln \big({{p_\svec(Y(x)[k(x)])
\overline{p_\svec(Y(x)[k(x)])}}\bigr)}}\cr &e^{i\int dx(\lambda_L(\svec(x)) +
V(Y(x))) }\cr}\ ,\eqnu$$ and that perturbation theory based on this expression
will respect the global structure of the configuration space.

An  analog of formula 5.2 is an alternative solution to the problem of
finding the phase
space path integral for quantum gravity. This quiestion has been addressed
in a large number of publications and often leads to measures of the form
$d\mu_{Liouville}$ and hence to configuration space measures
$d\mu_{invariant}$. It is then often argued that in dimensional
regularization the form of the measure is unimportant and hence that
finding the correct form of the measure is fruitless. As was clear from
the example in section three, this is not the case for the path
integral put forward here.
\vfill\eject

\leftline{\bf Acknowledgements}

We wish to thank Gary Gibbons, Stephen Hawking and Chris Isham
for helpful conversations. MJP  thanks for the Royal Society,
and NL thanks Trinity College, Cambridge for financial support.

\bigskip
\leftline{\bf References}
\bigskip
\item {[1]} M. Stone, {\it Linear Tranformations in Hilbert Space,}
Amer.Math.Soc.Colloq.Publ.15, (1932).

J. von Neumann, Ann. Math. {\bf 104}, (1931), 570.

\item {[2]} C. DeWitt-Morette, A. Maheshwari and B. Nelson,
Phys. Reps. {\bf 50}(1979) 255.

J. Glimm and A. Jaffe {\it Quantum Physics - A Functional Integral
Point of View},

Springer Verlag, New York (1981).

D.W. McLaughlin and L.S. Schulman, J. Math. Phys. {\bf 12}, (1971), 2520.

B.S. DeWitt, Rev. Mod. Phys. {\bf 29}, (1957), 377.

J.S. Dowker, J. Math. Phys. {\bf 17}, (1976), 1873.

\item{[3]} N. Linden and M.J. Perry, Nucl. Phys. {\bf B357}, (1991), 289.

\item{[4]} M. Pilati, Phys. Rev. {\bf D26}, (1982), 2645.

M. Pilati, Phys. Rev. {\bf D28}, (1983), 729.

C.J. Isham, Proc. R. Soc. London {\bf A351}, (1976), 209.

\item {[5]} N.P. Landsman and N. Linden, Nucl. Phys. {\bf B365}, (1991), 121.

\item {[6]} C.J. Isham, in Relativity, Groups and Topology 2 (Les Houches
           1983), B.S. DeWitt and R. Stora (eds.) North Holland, Amsterdam,
           (1984).

\item {[7]} B.S. DeWitt, Phys. Rev. {\bf 160}, (1967), 1113.

C.J. Isham and A.C. Kakas,  Class. Quantum Grav. {\bf 1}, (1984), 621.

\item{[8]} J.E. Roberts, Commun. Math. Phys. {\bf 3}, (1966), 98.

\item {[9]} R.P. Feynman and A.R. Hibbs, {\it Quantum Mechanics and Path
           Integrals} McGraw-Hill, New York, (1965).

\item {[10]} A. Terras, {\it Harmonic Analysis on Symmetric Spaces and
Applications 1}, Springer Verlag, New York, (1985).

A. Terras, {\it Harmonic Analysis on Symmetric Spaces and
Applications 2}, Springer

Verlag, New York, (1988).

\item {[11]} S. Helgason, {\it Groups and Geometric Analysis}, Academic Press,
New York, (1984).

\item {[12]}  A.O. Barut and R. Ra\c czka, {\it Theory of Group
Representations and Applications}, Polish Scientific Publishers, Warsaw,
(1977).

\item {[13]} S. Gindikin and F. Karpelevic, Sov. Math. Dokl. {\bf 3}, (1962),
962.

\item {[14]} L.D. Faddeev, Teor. Matem. Fiz. {\bf 1}, (1969), 3.

\item {[15]} M.A. Olshanetsky and A.M. Perelomov, Phys. Rep.
           {\bf 94}, (1983), 313.

\end